\documentclass[aps,prx,preprint,superscriptaddress,amsmath,amssymb]{revtex4-2}

\usepackage[version=4]{mhchem}
\usepackage{graphicx}
\usepackage[T1]{fontenc}
\usepackage[utf8]{inputenc}
\usepackage{siunitx}
\usepackage{xspace}
\usepackage{silence}
\usepackage{hyperref} 
\hypersetup{pdfborder=0 0 0,colorlinks=true,citecolor=blue,linkcolor=blue}

\DeclareUnicodeCharacter{200B}{{\hskip 0pt}}

\newcommand{\mapi}[0]{\texorpdfstring{\ce{MAPbI3}}{MAPbI3}\xspace}

\begin{document}


\title{Strain-Induced Tetragonal-to-Cubic Transition in \mapi from Machine-Learned Molecular Dynamics}

\author{Daan Holleman}
\affiliation{Materials to Optoelectronic Devices Group and Institute of Complex Molecular Systems, Eindhoven University of Technology, Eindhoven 5600 MB, The Netherlands}
\affiliation{MESA+ Institute for Nanotechnology, University of Twente, 7522 NB Enschede, The Netherlands}

\author{Luuk Klein Nagelvoort}
\affiliation{MESA+ Institute for Nanotechnology, University of Twente, 7522 NB Enschede, The Netherlands}

\author{Monica Morales-Masis}
\affiliation{Materials to Optoelectronic Devices Group and Institute of Complex Molecular Systems, Eindhoven University of Technology, Eindhoven 5600 MB, The Netherlands}

\author{Linn Leppert}
\email{l.leppert@bham.ac.uk}
\affiliation{School of Metallurgy and Materials, University of Birmingham, B15 2SE Birmingham, United Kingdom}
\affiliation{MESA+ Institute for Nanotechnology, University of Twente, 7522 NB Enschede, The Netherlands}

\date{\today}

\begin{abstract}
Metal-halide perovskites are promising photovoltaic absorbers whose optoelectronic properties are highly tunable. Epitaxial strain offers a route to stabilize desirable phases, tune band gaps, and reduce energetic disorder in these materials, but rational strain engineering requires predictive models. Zero-temperature ``strained-bulk'' calculations based on density functional theory (DFT) have been successful for oxide perovskites. However, these static calculations neglect dynamic disorder and the pronounced anharmonic effects that govern halide perovskites at ambient temperatures. Here, we demonstrate that molecular-dynamics simulations based on machine-learned force fields trained on-the-fly on first-principles DFT calculations reproduce the structural and optoelectronic response of methylammonium (MA) lead iodide (\mapi) to epitaxial strain at 300~K. We simulate both bulk and free-standing slab models of \mapi across a range of experimentally accessible biaxial strains by locking in-plane lattice vectors in nanosecond-long simulations. Our calculations predict a tetragonal-to-cubic phase transition at $\sim$0.5\% tensile strain, marked by a $c/a$ ratio of unity and the loss of the $a^0a^0c^-$ octahedral tilting pattern of the tetragonal phase in agreement with experiments on epitaxial \mapi films grown on KCl. MA cation orientations change with strain but retain preferential directions in the strain-induced cubic phase, unlike in its temperature-induced counterpart. Bulk and slab models yield consistent trends, with surface effects decaying within a few octahedral layers. Finally, we find a blue shift in band gap, consistent with experiment, when we compute the gap as an ensemble average over molecular dynamics snapshots, whereas static calculations on averaged structures predict a red shift. Our study positions molecular dynamics with constrained lattice vectors as a method for predicting finite-temperature strain phase diagrams and guiding the epitaxial design of perovskite absorbers.
\end{abstract}

\maketitle

\section{Introduction}

Metal-halide perovskites (MHPs) are among the most promising photovoltaic materials: their \ce{ABX3} composition allows extensive tuning of the optoelectronic properties, and their low-temperature, scalable processing has enabled rapid efficiency gains for solar-energy applications~\cite{jena_halide_2019, hautzinger_-site_2024, green_solar_2026}. For III-V semiconductors, epitaxy and strain engineering were key to new physics and large-scale commercial devices~\cite{capasso_band-gap_1987, dupuis_iiiv_2023}, and for oxide perovskites they have enabled functionalities absent in bulk \cite{schlom_strain_2007, haeni_room-temperature_2004}. Epitaxial strain can stabilize metastable phases and tune band gaps and other properties \cite{haeni_room-temperature_2004, langenberg_epitaxial_2015}, and phenomenological and first-principles methods have been used extensively to construct strain phase diagrams that predict these phases and their properties \cite{pertsev_effect_1998, dieguez_ab_2004, dieguez_first-principles_2005, rabe_theoretical_2005}. Achieving the same for MHPs has proven difficult. Compared with oxide perovskites, MHPs have soft lattices and are highly dynamic, with large structural distortions that strongly affect their optoelectronic properties~\cite{egger_what_2018, katan_entropy_2018, liang_structural_2023}. This makes them a far more challenging material class for epitaxial stabilization. Moreover, MHP films are typically solution-processed and polycrystalline, so that strain arises from thermal-expansion mismatch and processing, varies between grains, and is partly relaxed at grain boundaries~\cite{zhu_strain_2019, jones_lattice_2019, kim_importance_2020}, in contrast to the uniform, coherent biaxial strain imposed by epitaxy on a single-crystalline substrate. In addition, few lattice-matched substrates are available because MHP lattice parameters are large compared with those of conventional oxide and semiconductor substrates~\cite{shi_halide_2020, han_impact_2023, wang_research_2026}.

Despite these challenges, epitaxial growth of MHPs has recently been demonstrated, yielding films with improved stability and optoelectronic properties as well as control over the structural phase and device interfaces~\cite{wang_epitaxial_2019, chen_strain_2020, shi_halide_2020,solomon_room-temperature_2025, lu_layer-by-layer_2025, wei_perovskite_2025}. Chen \textit{et al.}~\cite{chen_strain_2020} grew $\alpha$-\ce{FAPbI3} thin films epitaxially on \ce{MAPbCl_xBr_{3-x}} single-crystals, whose halide composition tunes the lattice mismatch. The resulting strain tuned the band gap of the $\alpha$-\ce{FAPbI3} film and prolonged its phase stability, whereas unstrained films degraded to the photo-inactive $\delta$ phase. Lu \textit{et al.}\cite{lu_layer-by-layer_2025} used thermal evaporation for layer-by-layer heteroepitaxy of \ce{CsPbBr3} on \ce{PEA2PbBr4} single crystals, achieving angstrom-level thickness control and a lower Urbach energy (22~meV compared with 37~meV on quartz). Some of us demonstrated epitaxy of \mapi on KCl by pulsed laser deposition \cite{solomon_room-temperature_2025}. Structural characterization pointed to the formation of a cubic phase, and the photoluminescence band gap of 1.66~eV of 15-nm-thick films matched that of cubic \mapi measured at elevated temperature in spin-coated~\cite{milot_temperature-dependent_2015} and co-evaporated films~\cite{palazon_room-temperature_2019}, blue-shifted with respect to the 1.61~eV typical of tetragonal \mapi at room temperature~\cite{milot_temperature-dependent_2015}. These films also exhibited a remarkably low Urbach energy of 12.3~meV \cite{solomon_room-temperature_2025}.

These experiments raise questions that are difficult to answer experimentally. First, the tetragonal-to-cubic transition of \mapi is close to tricritical, and at room temperature the tetragonal distortion of bulk \mapi is small ($c/a \approx 1.009$)~\cite{whitfield_structures_2016}. Distinguishing a strain-stabilized cubic phase from a weakly tetragonal one is therefore challenging, particularly in polycrystalline films, and cubic \mapi has been reported at room temperature even in films grown without epitaxy~\cite{palazon_room-temperature_2019}. Second, diffraction probes the time- and space-averaged structure, so whether a strained film is cubic only on average or also locally remains open~\cite{morana_cubic_2023, beecher_direct_2016}. Third, the microscopic origin of the strain-induced blue shift and of the low energetic disorder is unclear. Answering these questions, and predicting which phases and properties are accessible by strain, requires a predictive theory of strained MHPs at finite temperature.

For oxide perovskites, static ``strained-bulk'' calculations, in which epitaxial strain is imposed by fixing the in-plane lattice parameters and relaxing the out-of-plane lattice parameter and internal coordinates, have been highly successful in predicting strain phase diagrams~\cite{dieguez_ab_2004, dieguez_first-principles_2005} and properties of strained phases such as the polarization~\cite{bungaro_epitaxially_2004, neaton_theory_2003}. The same approach has been applied to MHPs, predicting band-gap tunability of \ce{CsSnI3} and \ce{CsPbI3} for biaxial strains of up to $\pm 3\%$~\cite{grote_strain_2015} and strain-induced polar phases with sizable Rashba splitting in \ce{CsPbI3} and \mapi~\cite{leppert_electric_2016}. For \mapi on KCl, however, strained-bulk calculations of the tetragonal phase yielded a cubic unit cell with substantial octahedral tilting, at odds with the cubic phase, and band gaps of the cubic phase could only be obtained by constraining Pb and I to their high-symmetry positions~\cite{solomon_room-temperature_2025}. The tetragonal-to-cubic transition and its effect on the band gap thus remained inaccessible.

The origin of this failure is that the tetragonal and cubic phases of MHPs are dynamically unstable at 0~K and stabilized only at finite temperature, through anharmonic double-well octahedral tilting and librational motion of the A-site cations~\cite{lahnsteiner_finite-temperature_2018, wiktor_quantifying_2023}. Local tilting characteristic of the tetragonal phase persists above the tetragonal-to-cubic transition~\cite{beecher_direct_2016, wiktor_quantifying_2023}, and the cubic phase is better described as an average over locally distorted structures~\cite{zhao_polymorphous_2020}. The special displacement method, including its extension to anharmonic lattices~\cite{zacharias_one-shot_2016, zacharias_theory_2020, zacharias_anharmonic_2023} and polymorphous supercells~\cite{zhao_polymorphous_2020}, provide access to finite-temperature structures and band gaps of MHPs from static calculations. Molecular dynamics (MD) offers a complementary route that requires no phonon-based reference structure: it samples large-amplitude anharmonic motion, including A-site cation reorientation, directly, and allows phase transitions to emerge as a function of an external control parameter. MD simulations reproduce the phase transitions of \mapi~\cite{lahnsteiner_finite-temperature_2018, jinnouchi_phase_2019} and the temperature dependence of its band gap~\cite{quarti_structural_2016, wiktor_quantifying_2023}, and machine-learned force fields trained on-the-fly on DFT now make nanosecond simulations of large supercells feasible~\cite{jinnouchi_--fly_2019, lahnsteiner_anharmonic_2022}. An analogous finite-temperature description of epitaxial strain in MHPs, however, has been lacking.

Here, we use MD simulations based on machine-learned force fields trained on-the-fly on first-principles DFT calculations to model epitaxially strained \mapi at 300~K. We impose biaxial strain by fixing the in-plane lattice vectors and simulate both a bulk and a free-standing slab model of a thin film on nanosecond timescales. Both models yield consistent structural trends, indicating that free surfaces do not significantly alter the strain response. With increasing tensile strain, the $c/a$ ratio decreases continuously and reaches unity at approximately 0.5\% strain, where the $a^0a^0c^-$ tilting pattern of the tetragonal phase gives way to an $a^0a^0a^0$ pattern. The tetragonal-to-cubic phase transition is gradual: short-range tilt correlations persist near the transition, so that the strain-induced cubic phase, like its thermally induced counterpart, is cubic only on average. Unlike in the high-temperature cubic phase, however, the \ce{MA+} cations retain preferential orientations, which we attribute to the insufficient thermal energy at 300~K to overcome the barriers between them. The most isotropic distribution occurs near 0.5\% strain, coinciding with the low Urbach energy observed experimentally. Band gaps computed as ensemble averages over MD snapshots blue-shift under tensile strain, consistent with experiment, whereas static calculations on the time-averaged structures predict the opposite trend near the transition, because the band gap depends non-linearly on the instantaneous octahedral tilts. Our results establish finite-temperature MD with constrained lattice vectors as a route toward strain phase diagrams of MHPs, analogous to those that have guided strain engineering of oxide perovskites.

\section{Methods}

\begin{figure*}
    \centering
    \includegraphics[scale=1,page=1]{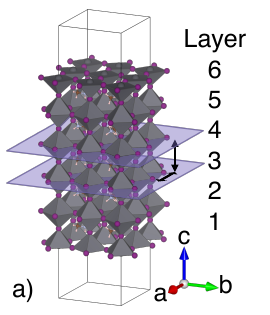}
    \includegraphics[scale=1,page=2]{figure1.pdf}
    \includegraphics[scale=1,page=3]{figure1.pdf}
    \caption[Definitions of the structural parameters of \mapi.]{Structural quantities used to characterize the bulk and slab models of \mapi. (a) Slab model with octahedral layers numbered 1-6; the out-of-plane lattice parameter $c$ is the distance between the Pb planes of the two central layers (3 and 4, shaded), from which the pseudocubic $c/a$ ratio is obtained. (b) Octahedral rotation angle about the $c$ axis, viewed along $c$: the azimuthal angle of an equatorial iodide relative to the time-averaged Pb-I-Pb bond direction. (c) Orientation of the \ce{MA+} cation, defined by the C-N vector, with polar angle $\theta$ relative to $c$ and azimuthal angle $\phi$ in the $ab$ plane.}
    \label{fig:structural parameters}
\end{figure*}

All calculations are performed using the Vienna Ab-initio Simulation Package (VASP)~\cite{kresse_ab_1993, kresse_efficient_1996, kresse_efficiency_1996}. We use the strongly constrained and appropriately normed (SCAN) meta-generalized gradient approximation~\cite{sun_strongly_2015} in the MD simulations, which has been shown to yield good agreement with accurate many-body calculations~\cite{bokdam_assessing_2017} and the experimental structural properties of \mapi between 250~K and 400~K~\cite{lahnsteiner_finite-temperature_2018}. To enable efficient finite-temperature simulations on nanosecond timescales, we use on-the-fly training of machine-learned force fields (MLFFs) based on forces from DFT~\cite{jinnouchi_--fly_2019, jinnouchi_phase_2019, lahnsteiner_anharmonic_2022}. Details of the MLFF training are described in the Supplementary Material (SM) \cite{SM}. We use the projector augmented wave (PAW) method~\cite{blochl_projector_1994} with valence electronic configurations ($6s^26p^2$) for Pb, ($5s^2 5p^5$) for I, ($2s^22p^2$) for C, ($2s^22p^3$) for N, and ($1s^1$) for H. From convergence tests (Figure~S1), we determine that an energy cutoff of 450~eV and $\Gamma$-centered $\mathbf{k}$-point grids of $2 \times 2 \times 2$ for the bulk and $2 \times 2 \times 1$ for the slab result in converged total energies to within 1~meV/atom. Band gap calculations use the same energy cutoff and $\mathbf{k}$-point sampling but the $r^2$SCAN functional for numerical stability~\cite{furness_accurate_2020}. Spin-orbit coupling is not included as it has negligible effects on the structure \cite{Egger2014}. All MD simulations are performed at 300~K using the Langevin thermostat in an $NpT$ ensemble. The bulk simulations use a time step of 0.5~fs, equilibration time of 300~ps, and simulation time of 5~ns. The slab simulations use a time step of 1.5~fs, equilibration time of 300~ps, and simulation time of 13~ns. Further details on convergence tests and accuracy of the machine-learned force-fields can be found in the SM and Figure~S4.

To assess whether the free surfaces of thin films alter their response to strain, we compare a bulk model with a free-standing slab model of \mapi. Both models impose coherent biaxial strain throughout the structure and therefore describe films below the critical thickness for strain relaxation. Relaxation via misfit dislocations is beyond the scope of our models. Experimentally, epitaxial \mapi films on KCl are fully strained at a thickness of 15~nm, whereas thicker films (up to 70~nm) show partial out-of-plane relaxation~\cite{solomon_room-temperature_2025}. We therefore compare our results to the 15-nm-thick films. We note that a 24-layer slab, which we use to test convergence with slab thickness (Fig.~S3), corresponds to a thickness of approximately 15~nm, comparable to these films.

We model bulk \mapi using a $2\times2\times2$ supercell of the pseudocubic unit cell and the thin film using a $2\times2\times6$ slab with symmetric (001) \ce{PbI2} termination on both sides~\cite{geng_effect_2015, lodeiro_methodological_2020}, centered in the simulation cell with 20~\AA\ of vacuum to prevent interactions between periodic images (Fig.~S2). Comparing this slab with a $2\times2\times24$ slab, we find that the out-of-plane Pb-I bond lengths converge within two octahedral layers of the surface (Fig.~S3). Starting all simulations from the cubic structure ensures that no tilting pattern or tetragonal distortion is imposed. We use a $2\times2$ in-plane cell, which is commensurate with the in-phase and out-of-phase tilting patterns relevant here. This cell size does, however, constrain tilt correlations to the nearest-neighbor level in plane and limits configurational averaging of band gaps. We therefore focus on strain-induced trends, and discuss the consequences of the finite cell size where relevant.

Because SCAN slightly overestimates the lattice parameters of \mapi~\cite{xue_first-principles_2021}, we define strain relative to the in-plane lattice parameter at which the pseudocubic $c/a$ ratio equals its experimental room-temperature value of 1.009~\cite{whitfield_structures_2016}. This allows trends with strain to be compared quantitatively with experiment.

We characterize the strained structures by three quantities (Fig.~\ref{fig:structural parameters}). For the slab, we use only the two central octahedral layers (layers 3 and 4) while for the bulk, we use all atoms. First, the pseudocubic $c/a$ ratio is obtained from the time-averaged vertical distance between the Pb planes of two adjacent layers, $c$, and the imposed in-plane pseudocubic lattice parameter $a$ (Fig.~\ref{fig:structural parameters}a). Second, the octahedral rotation angle about the $c$ axis is the azimuthal angle of each equatorial iodide in cylindrical coordinates centered on the time-averaged position of its Pb atom, measured relative to the time-averaged Pb-I-Pb bond direction (Fig.~\ref{fig:structural parameters}b). A single rotation angle suffices because the tetragonal ($a^0a^0c^-$) and cubic ($a^0a^0a^0$) phases have no tilts about the in-plane axes~\cite{glazer_classification_1972, lahnsteiner_finite-temperature_2018, wiktor_quantifying_2023}. We verify this a posteriori (Fig.~S9). Third, the orientation of each \ce{MA+} cation is defined by its C$\rightarrow$N vector, which also defines the molecular dipole~\cite{leguy_dynamics_2015, lahnsteiner_finite-temperature_2018}, expressed by its polar angle $\theta$ with respect to the $c$ axis and its azimuthal angle $\phi$ in the strain plane (Fig.~\ref{fig:structural parameters}c).

\section{Results and Discussion}

\begin{figure*}[ht]
    \centering
    \begin{minipage}[t]{0.495\linewidth}
        \centering
        \includegraphics[width=\linewidth,page=1]{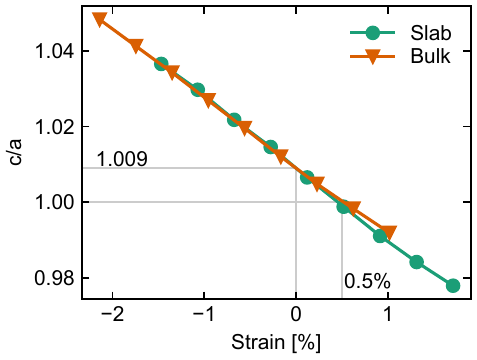}\\
        (a)
    \end{minipage}
    \hfill
    \begin{minipage}[t]{0.495\linewidth}
        \centering
        \includegraphics[width=\linewidth,page=2]{figure2.pdf}\\
        (b)
    \end{minipage}
    \caption[Structural parameters as a function of biaxial strain.]{Structural response of \mapi to biaxial strain from MD simulations of the bulk and slab models. (a) Pseudocubic $c/a$ ratio, for the slab evaluated in the two central layers. For each model, 0\% strain is defined by $c/a = 1.009$, the experimental room-temperature value~\cite{whitfield_structures_2016}. (b) Distributions of the octahedral rotation angle about the $c$ axis (violin plots); lines connect the modes of the distributions as a guide to the eye. The arrow at 0\% strain marks the experimental rotation angle of \ang{15}~\cite{schuck_influence_2019, schuck_thermal_2024}.}
    \label{fig:result comparison}
\end{figure*}

Fig.~\ref{fig:result comparison}a shows that the $c/a$ ratio decreases continuously with increasing tensile strain: the structure contracts out of plane as it expands in plane. In both models, $c/a$ reaches unity at approximately 0.5\% tensile strain, in agreement with experimental results of epitaxial \mapi on KCl~\cite{solomon_room-temperature_2025}. Block averaging yields a standard error of the mean of at most 0.0003 (Fig.s~S5 and S6)~\cite{flyvbjerg_error_1989}.

At identical in-plane lattice parameters, the slab and bulk models yield different $c/a$ ratios, so that the in-plane lattice parameter corresponding to 0\% strain differs by 0.7\% between them. This offset reflects the effect of the free surfaces of the slab. Once each model is referenced to its own unstrained state, however, both follow the same trend with strain, indicating that the surfaces shift the reference structure but do not alter the strain response of the slab interior. This is consistent with the rapid convergence of the out-of-plane Pb-I bond lengths within two octahedral layers of the surface (Fig.~S3), which we attribute to the softness of the Pb-I framework.

Fig.~\ref{fig:result comparison}b shows the distributions of the octahedral rotation angle sampled along the MD trajectories at each strain. We take the mode of each distribution as the characteristic rotation angle at that strain. For compressive strain, there are two well separated modes which merge into one with increasing tensile strain. Experimentally, the octahedral rotation angle at room temperature is \ang{15}~\cite{schuck_thermal_2024, schuck_influence_2019} indicated in Fig.~\ref{fig:result comparison}b by the arrow at 0\% strain. We find our result for the slab model to be in good agreement with this value but the bulk model underestimates this angle by approximately \ang{7}. Note that near 0\% strain the angle distributions are broad and their modes therefore carry a large uncertainty, so that the difference between the slab and bulk results is not conclusive.

With increasing tensile strain, the modal rotation angle decreases and vanishes at approximately 0.5\% strain in the slab and already at 0.2\% in the bulk. Together with $c/a = 1$, this marks the onset of a cubic phase of \mapi. For the three largest tensile strains, where the distributions are single-peaked, block averaging yields a standard error of the mean of at most \ang{0.1} (Figs.~S7 and S8).

The distributions shown in Fig.~\ref{fig:result comparison}b reveal separate modes of the rotation angle, but do not establish a long-range tilting pattern, which requires the rotations of neighboring octahedra to be correlated~\cite{glazer_classification_1972}. We therefore analyze the correlation between the rotation angles of octahedra in adjacent layers along $c$. A positive correlation corresponds to in-phase rotation ($a^0a^0c^+$) and a negative correlation to out-of-phase rotation ($a^0a^0c^-$). In the absence of any correlation, the rotations average to zero and the structure adopts the $a^0a^0a^0$ pattern of the cubic phase on average, even though instantaneous rotations remain finite.

\begin{figure*}
    \centering
    \includegraphics[scale=1,page=2]{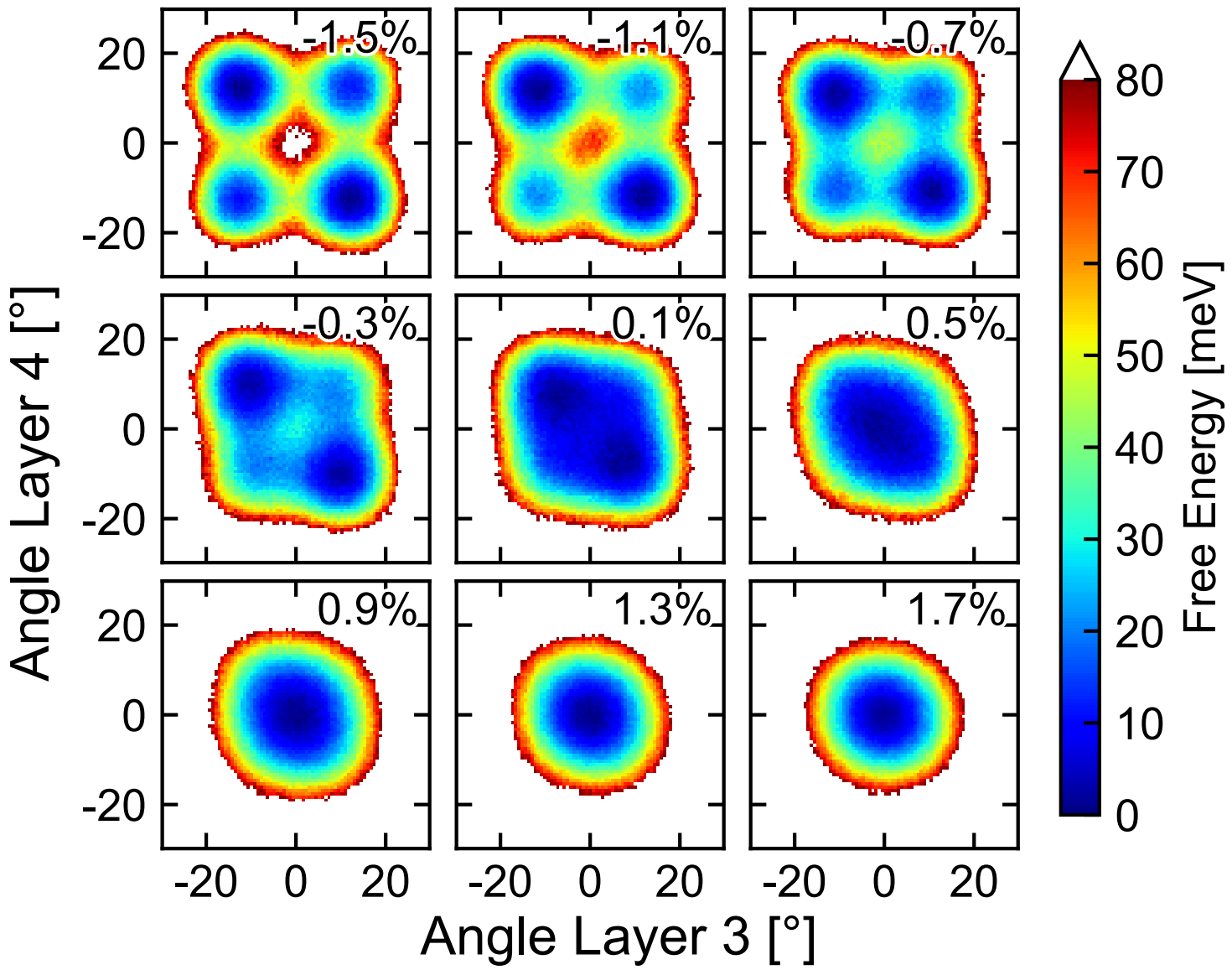}
    \caption[Effective free energy surfaces of the octahedral tilting.]{Effective free-energy surfaces of octahedral rotations in adjacent layers of the \mapi slab at different strains. Each panel shows the free energy, obtained by Boltzmann inversion of the joint probability distribution, as a function of the rotation angles of vertically stacked iodides in the two central layers (layers 3 and 4). Minima in the upper-left and lower-right quadrants correspond to out-of-phase ($a^0a^0c^-$), those in the other two quadrants to in-phase ($a^0a^0c^+$) rotations.}
    \label{fig:result angle correlation}
\end{figure*}

We quantify this correlation by constructing, for each MD frame, a two-dimensional histogram of the rotation angles of pairs of iodides in adjacent layers along $c$. Boltzmann inversion of the resulting joint probability distribution yields an effective free-energy surface~\cite{lahnsteiner_finite-temperature_2018}, shown in Fig.~\ref{fig:result angle correlation} for the two central layers of the slab at different strains. Under compressive strain, the free-energy surface exhibits two deep minima in the upper-left and lower-right quadrants, where the rotation angles of adjacent layers have opposite signs, characteristic of out-of-phase ($a^0a^0c^-$) tilting. The two minima corresponding to in-phase tilting lie 10~meV higher at $-1.5\%$ strain, and the minima are separated by barriers of 45~meV, approximately $1.7\,k_\mathrm{B}T$ at 300~K. As the strain changes from compressive to tensile, these barriers vanish. At 0.5\% tensile strain, where the barriers have disappeared, the free-energy surface nevertheless remains elongated along the out-of-phase diagonal, indicating that out-of-phase correlations between adjacent layers persist without long-range order. Similar short-range order has been reported above the thermal tetragonal-to-cubic transition: local tilting characteristic of the tetragonal phase persists in cubic halide perovskites~\cite{wiktor_quantifying_2023, beecher_direct_2016}, and the cubic phase is better described as an average over locally distorted structures than as the ideal high-symmetry structure~\cite{zhao_polymorphous_2020}. At larger tensile strains, the free-energy surfaces become circular, i.e., the rotation angles of adjacent layers are uncorrelated. Together with the equivalent analysis of in-plane rotations (Fig.~S9), this shows that the structure adopts the $a^0a^0a^0$ pattern on average.

The orientations of the \ce{MA+} cations also change with strain. In unstrained structures, the orientational distribution of the cations is nearly isotropic in the high-temperature cubic phase but restricted to a few preferred directions at lower temperatures in the tetragonal and frozen in the orthorhombic phase~\cite{jinnouchi_phase_2019}. Cation orientation and the inorganic framework are coupled~\cite{li_atomic_2016}, and the increase in orientational isotropy with temperature reflects, at least in part, the thermal energy available to the cations to overcome the energy barriers between orientations. Because our strain series is performed at constant temperature, this contribution is absent, raising the question of whether strain alone can restore the isotropic cation orientations of the high-temperature cubic phase.

\begin{figure*}[ht]
    \centering
    \includegraphics[scale=1]{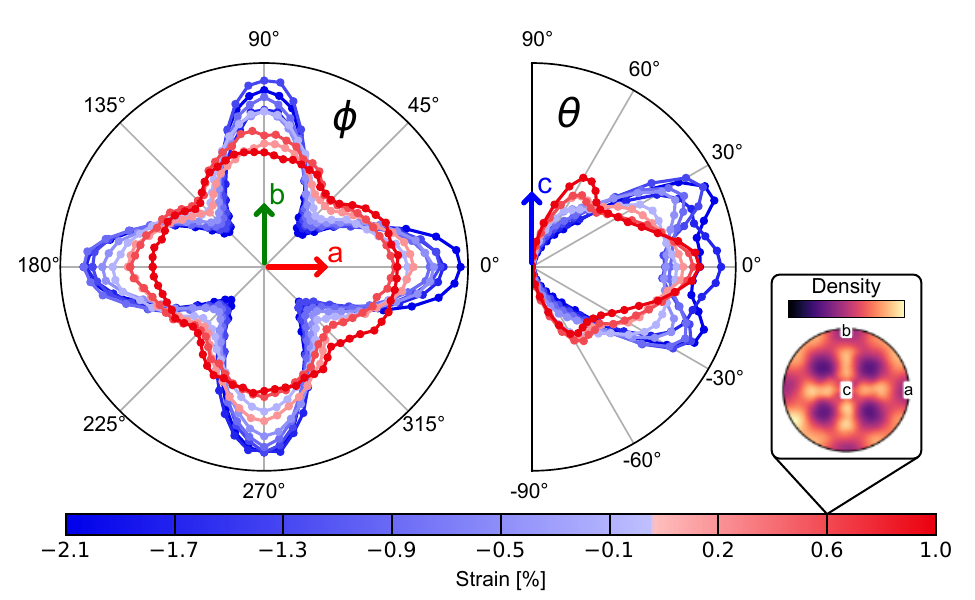}
    \caption[Polar orientations of the \ce{MA+} cations.]{Histograms of the polar orientations of the \ce{MA+} cations as a function of strain. The left panel shows the spherical coordinate $\phi$ which corresponds to directions within the strain plane. The right panel shows the spherical coordinate $\theta$ which is normal to the biaxial strain plane. The inset at 0.6\% strain shows a pole plot in which the occurrences of different orientations are shown by the color.}
    \label{fig:result ma orientations}
\end{figure*}

Fig.~\ref{fig:result ma orientations} shows the distributions of the \ce{MA+} orientation angles as a function of strain. At all strains, the cations retain preferred orientations, and the azimuthal distribution is fourfold symmetric in the strain plane. The preference is most pronounced under compressive strain, where the cations predominantly point to $\pm \ang{30}$ along the $ab$ lattice vectors, and are much less likely to lie in-plane and diagonal to the $ab$ lattice vectors. This preference weakens toward 0\% strain, and the distribution is most isotropic near 0.5\% tensile strain. The pole figure at 0.6\% strain (see Fig.~S11 for other strains), the simulation closest to a cubic cell, shows threefold symmetry within each octant, consistent with the local cubic symmetry of the octahedral framework. At larger tensile strain, the cell flattens and this symmetry is lost. Strain therefore does not restore the isotropic cation orientations of the high-temperature cubic phase, but yields a distribution with the full symmetry of the cubic lattice. Since isotropic A-site cation dynamics have been linked to low Urbach energies and high photoluminescence quantum yields~\cite{dubajic_dynamic_2025, dubajic_thermal_2026}, this is consistent with the low Urbach energy of 12.3~meV measured for 15-nm-thick epitaxial films~\cite{solomon_room-temperature_2025}.

\begin{figure}
    \centering
    \includegraphics[scale=1,page=1]{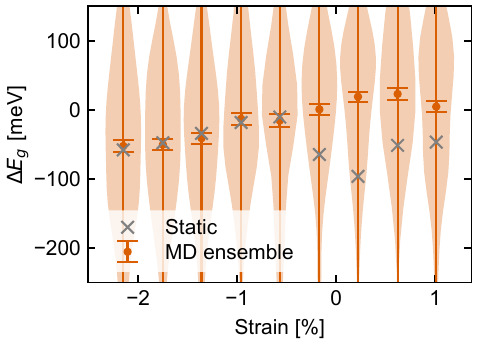}
    \caption[Change in band gap of the bulk structure with strain.]{Strain dependence of the band gap of the bulk model. Violin plots show the distributions of band gaps computed for 200 snapshots per strain sampled from the MD trajectories; circles with error bars (MD ensemble) give their means and standard errors of the mean. Gray crosses show band gaps of static structures constructed from the modal $c/a$ ratio and rotation angle. Band gaps are given relative to the 0\% intercept of a linear fit to the MD-ensemble means.}
    \label{fig:bandgap}
\end{figure}

Finally, we examine how strain affects the band gap, comparing static calculations on time-averaged structures with ensemble averages over MD snapshots. We restrict this analysis to the bulk model, which responds to strain in the same way as the slab interior and avoids surface states. For the static calculations, we construct one structure per strain from the modal values of $c/a$ and the octahedral rotation angle  (Fig.~\ref{fig:result comparison}a and b). The \ce{MA+} cations are kept fixed in an arrangement with zero net dipole moment, as they do not contribute electronically to the band edges~\cite{brivio_structural_2013}. The resulting band gaps are shown as gray crosses in Fig.~\ref{fig:bandgap}. Going from compressive to tensile strain, the band gap decreases, because the vanishing octahedral rotation straightens the Pb-I-Pb bonds and increases the overlap between Pb and I orbitals~\cite{filip_steric_2014}. For higher tensile strain, the rotation angle remains zero, but the increasing tetragonal distortion lengthens the in-plane Pb-I bonds, reducing the orbital overlap and increasing the band gap. These static calculations differ from those of Ref.~\citenum{solomon_room-temperature_2025}, in which the cubic phase was enforced at all strains by fixing the Pb and I positions. There, the band gap increased monotonically with tensile strain, consistent with our static results beyond the transition.

The static calculations thus predict a red shift across the transition, opposite to the blue shift observed experimentally between tetragonal and epitaxially strained cubic \mapi. Ensemble-averaged band gaps, in contrast, recover the experimental trend: compressive strain red-shifts and tensile strain blue-shifts the band gap (Fig.~\ref{fig:bandgap}). Analogous to approaches used for the temperature dependence of the band gap~\cite{quarti_structural_2016, wiktor_quantifying_2023, seidl_anharmonic_2023, zacharias_anharmonic_2023}, we compute the band gap for 200 snapshots per strain, sampled every 11~ps from the bulk MD trajectories, which directly sample the anharmonic ensemble including \ce{MA+} disorder. This interval exceeds the correlation times of both \ce{MA+} reorientation~\cite{mattoni_methylammonium_2015, govinda_behavior_2017} and octahedral rotations, so that the snapshots are statistically independent. The band-gap distributions are broad, with standard deviations of 98-130~meV, but their means, with standard errors of 7-9~meV, show a clear blue shift under tensile strain. The difference from the static results arises because the instantaneous rotation angles remain finite even where the modal angle vanishes: since the band gap depends non-linearly on the rotation angle, the average band gap differs from the band gap of the average structure.

At 0.5\% tensile strain, the calculated blue shift is 12~meV, smaller than the experimental difference of 50~meV from photoluminescence~\cite{milot_temperature-dependent_2015, solomon_room-temperature_2025}. Since Stokes shift and excitonic effects are small in \mapi at room temperature, the discrepancy most likely reflects limitations of our calculations. First, we neglect spin-orbit coupling and many-body corrections, which can modify the strain dependence of the band gap, particularly through the Pb-$p$-derived conduction band. Second, the $2\times2\times2$ supercell limits the configurational averaging of band gaps, and larger supercells may change the magnitude of the shift. Third, our unstrained reference, defined by $c/a = 1.009$, may not correspond exactly to the state of the reference films in experiment.

\section{Conclusions}

In summary, we have shown that MD simulations based on machine-learned force fields with constrained in-plane lattice vectors capture the structural and optoelectronic response of \mapi to epitaxial strain at room temperature. With increasing tensile strain, $c/a$ decreases continuously and reaches unity at approximately 0.5\% strain, where the $a^0a^0c^-$ tilting pattern gives way to $a^0a^0a^0$ on average, in agreement with experiments on epitaxial \mapi films on KCl~\cite{solomon_room-temperature_2025}. Ensemble-averaged band gaps reproduce the experimentally observed blue shift, whereas static calculations on time-averaged structures predict the opposite trend across the transition. As for the temperature dependence of the band gap~\cite{quarti_structural_2016, wiktor_quantifying_2023}, finite-temperature fluctuations must therefore be included to describe the strain response of \mapi. The strain-induced cubic phase resembles its thermally induced counterpart in being cubic only on average, with short-range tilt correlations persisting near the transition~\cite{beecher_direct_2016, wiktor_quantifying_2023, zhao_polymorphous_2020}. Unlike the high-temperature phase, however, it retains preferred \ce{MA+} orientations, suggesting that the thermal energy at 300~K is insufficient for the cations to overcome the barriers between orientations. Bulk and slab models show the same strain response, indicating that free surfaces shift the reference structure but do not alter the strain response of thin films.

The continuous character of the transition has implications for the interpretation of experiments. Because unstrained \mapi is only weakly tetragonal at room temperature, small strains, whether epitaxial or arising from thermal mismatch or processing, may suffice to render it cubic on average, and a strain-stabilized cubic phase may be difficult to distinguish from a weakly tetragonal one, particularly in polycrystalline films. Structural probes sensitive to local order, combined with simulations such as ours, can help resolve such cases.

Several aspects of our approach leave room for refinement. The calculated blue shift of 12~meV at 0.5\% strain is smaller than the experimental shift, which may be due to the neglect of spin-orbit coupling and many-body corrections, the limited in-plane supercell size, and differences between our unstrained reference and the experimental reference films. Our models impose coherent strain throughout and therefore do not capture strain relaxation in thicker films, and the free-standing slab neglects the perovskite-substrate interface, which may become relevant for the thinnest films or for chemically interacting substrates. Larger supercells, explicit substrate models, and band-gap calculations beyond DFT would address these limitations.

More broadly, our results demonstrate that finite-temperature MD with constrained lattice vectors can describe strain-driven phase transitions in MHPs, in close analogy to established approaches for temperature-driven transitions. Extending this approach to other compositions, temperatures, and dimensionalities would enable finite-temperature strain phase diagrams of MHPs, similar to those that have guided strain engineering of oxide perovskites~\cite{dieguez_ab_2004, dieguez_first-principles_2005}, and thereby support the rational design of strain-engineered perovskite absorbers.

\begin{acknowledgments}
    This research was performed at Eindhoven University of Technology and the University of Twente, as part of SolarLab within SolarNL, a national research, innovation and industrial development program funded by the Netherlands National Growth Fund. We acknowledge computational resources provided by the Dutch National Supercomputing Center Snellius supported by the SURF cooperative.
\end{acknowledgments}

%

\end{document}